\documentclass{egpubl}
\usepackage{eurovis2020}
\SpecialIssuePaper         

\usepackage[T1]{fontenc}
\usepackage{dfadobe}  
\usepackage{microtype}

\usepackage{cite}  
\BibtexOrBiblatex
\electronicVersion
\PrintedOrElectronic
\ifpdf \usepackage[pdftex]{graphicx} \pdfcompresslevel=9
\else \usepackage[dvips]{graphicx} \fi

\usepackage{egweblnk}

\graphicspath{{figures/}{pictures/}{images/}{./}} 

\usepackage{amsmath}
\usepackage{amssymb}
\usepackage{pifont,txfonts}
\usepackage{color}
\usepackage{colortbl}
\usepackage{xspace}
\usepackage{wrapfig}
\usepackage{float}
\usepackage{balance}
\usepackage{dblfloatfix}

\newcommand{\etal}{\emph{et al.}\xspace}
\newcommand{\mypar}[1]{\smallskip\noindent{\bfseries #1.}}
\definecolor{Gray}{gray}{0.7}
\definecolor{babyblue}{rgb}{0.54, 0.81, 0.94}

\title{Quantitative Comparison of Time-Dependent Treemaps}

\author[E. Vernier, M. Sondag, J. Comba, B. Speckmann, A. Telea, and K. Verbeek]
{\parbox{\textwidth}{\centering Eduardo Vernier$^2$\orcid{0000-0001-9483-4978}, Max Sondag$^1$\orcid{0000-0003-3309-638X}, Jo\~ao Comba$^2$\orcid{0000-0003-2921-2130}, Bettina Speckmann$^1$\orcid{0000-0002-8514-7858}, Alexandru Telea$^3$\orcid{0000-0003-0750-0502}, and Kevin Verbeek$^1$\orcid{0000-0003-3052-4844}
    }
        \\
{\parbox{\textwidth}{\centering $^1$TU Eindhoven , Netherlands\\
         $^2$UFRGS, Brazil\\
		 $^3$Utrecht University, Netherlands
       }
}
}

\begin{document}


\maketitle
\begin{abstract}
Rectangular treemaps are often the method of choice to visualize large hierarchical datasets. Nowadays such datasets are available over time, hence there is a need for (a) treemaps that can handle time-dependent data, and (b) corresponding quality criteria that cover both a treemap's visual quality and its stability over time.
In recent years a wide variety of (stable) treemapping algorithms has been proposed, with various advantages and limitations. 
We aim to provide insights to researchers and practitioners to allow them to make an informed choice when selecting a treemapping algorithm for specific applications and data. To this end, we perform an extensive quantitative evaluation of rectangular treemaps for time-dependent data.
As part of this evaluation we propose a novel classification scheme for time-dependent datasets. Specifically, we observe that the performance of treemapping algorithms depends on the characteristics of the datasets used. We identify four potential representative features that characterize time-dependent hierarchical datasets and classify all datasets used in our experiments accordingly. We experimentally test the validity of this classification on more than 2000 datasets, and analyze the relative performance of 14 state-of-the-art rectangular treemapping algorithms across varying features. Finally, we visually summarize our results with respect to both visual quality and stability to aid users in making an informed choice among treemapping algorithms.
All datasets, metrics, and algorithms are openly available to facilitate reuse and further comparative studies.


\end{abstract}  
\section{Introduction}
\label{sec:introduction}

Treemaps are one of the best-known methods for visualizing large hierarchical datasets. Given an input tree whose leaves have several attributes, treemaps recursively partition a 2D spatial region into cells whose visual attributes (area, color, shading, or annotation) encode the tree's data attributes. Compared to other methods such as node-link techniques, treemaps effectively use all available screen pixels to show data, and thus can display trees of tens of thousands of nodes on a single screen. Most treemaps use rectangles, although there are alternative models such as Voronoi treemaps~\cite{balzer05b}, orthoconvex and L-shaped treemaps~\cite{deberg14}, and Jigsaw treemaps~\cite{jigsaw}. 
In this paper, we focus exclusively on rectangular treemaps.

The input for a rectangular treemap is a rectangle $R$ and a set of non-negative values $a_1, \ldots, a_n$ together with a hierarchy on these values (represented by a tree). The output is a treemap $T$, which is a recursive partition of $R$ into a set $\mathcal{R}=\{R_1, \ldots, R_n \}$ of interior-disjoint rectangles, where $(a)$ each rectangle $R_i$ has area $a_i$, and $(b)$ the regions of the children of an interior node of the hierarchy form a rectangle (associated with their parent). Such a partition of a rectangle into a set of disjoint rectangles is also called a \emph{rectangular layout}, or \emph{layout} for short. Typically the input values are \emph{normalized}, that is, the sum $A = \sum_i a_i$ corresponds to the area of $R$.

Nowadays, large hierarchical datasets are also available over time. Hence, there is a need for \emph{time-dependent} treemaps which display changing trees and data values. Ideally, such time-dependent treemaps enable the user to easily follow structural changes in the tree and in the data. In a time-dependent setting, the input values become functions $a_i\colon [0, X] \rightarrow \mathbb{R}_{\geq 0}$ for each $i$, where the discrete domain $[0, X]$ represents the different time steps in the data. We assume that the hierarchy on the values and $R$ are not time-dependent, and that the values $a_i$ are properly normalized for each time step separately. We use the special value $a_i(t) = 0$ to represent that data element $i$ is not present at time $t$; and we speak of insertions or deletions if $a_i(t)$ starts or stops to be nonzero, respectively.

The \emph{visual quality} of rectangular treemaps is usually measured via the aspect ratio of its rectangles. This indicator can become arbitrarily bad: Consider a treemap that consists of only two rectangles. If the area of one of these rectangles tends towards zero, then its aspect ratio tends towards infinity. Nagamochi and Abe~\cite{nagamochi2007approximation} describe an algorithm (APP) which computes, for a given set of values and a hierarchy, a treemap which provably approximates the optimal aspect ratio. De Berg~\etal~\cite{deberg14} prove that minimizing the aspect ratio for rectangular treemaps is strongly NP-complete. Kong~\etal~\cite{Kong2010} propose perceptional guidelines to improve treemap design and Zhou~\etal~\cite{Zhou2017} perform user studies to test the effectiveness of different rectangular treemapping algorithms. Recently Lu~\etal~\cite{lu2017golden} argue that the optimal aspect ratio for treemaps should, in fact, be the golden ratio. In Section~\ref{sec:algorithms}, we describe the state-of-the-art of rectangular treemaps in detail along with the various characteristics of rectangular treemaps.

For time-dependent treemaps, a second quality criterion is \emph{stability}. Ideally, small changes in the data should result only in small changes in the treemap. Such stable behavior ensures that the only changes the user sees are due to the data, and not due to the decisions the algorithm makes.
In recent years a few non-rectangular treemaps were specifically developed for time-dependent data. Hahn~\etal~\cite{hahn10} and Hees and Hage~\cite{hees17} describe stable versions of Voronoi treemaps. Chen~\etal~\cite{Chen2017} propose a small-multiple metaphor to visualize time-dependent hierarchies. Their algorithm computes a global layout for all time steps simultaneously, but does not handle insertions or deletions. Scheibel~\etal~\cite{Scheibel2018} give an algorithm that maps changes in the data onto an initial layout. However, ``treemaps'' of subsequent time steps are not proper rectangular layouts as white space is introduced when resolving overlaps between rectangles.
Finally, Lukasczyk~\etal~\cite{lukasczyk2017nested} and K{\"o}pp and Weinkauf~\cite{kopp2019temporal} show how to compute static overviews of the whole evolution of time-dependent hierarchical data sets, while
Guerra-G{\'o}mez~\etal~\cite{StemView} and Card~\etal~\cite{TimeTree} present interactive visualization tools.

\mypar{Contribution} Despite their enduring popularity, a comprehensive evaluation of treemaps is currently lacking, even more so for the time-dependent case. Individual papers tend to report on only a few algorithms and evaluate only a few datasets, often without a principled discussion of quality metrics. To provide insights to both researchers and practitioners and to allow them to make an informed choice when selecting a treemap for their specific application and data, we perform an extensive quantitative evaluation of rectangular treemaps for time-dependent data. Our three main contributions are:

\noindent
\textbf{(1)} We introduce a new method to measure the stability of time-dependent treemaps which explicitly considers the input data (Section~\ref{sec:stablity}). An algorithm is stable if small changes in the input data result in small changes in the layout, that is, data change and layout change correlate positively. Previously proposed stability metrics measure only the layout change and conclude that small layout changes are a sign of a stable algorithm. However, to properly measure stability, we also need to capture the data change and then correlate data and layout change. Here, we have to
overcome the difficulty that the data and the layout space are a priori incomparable. We solve this problem by introducing the concept of a \emph{baseline treemap} $T^*$ which represents the minimum amount of change that any time-dependent treemap must incur (given the input data) when moving from treemap $T$ to the next treemap $T'$.

\noindent
\textbf{(2)} We propose a novel classification scheme for time-dependent datasets. Specifically, based on our discussion of the state-of-the-art of treemaps in Section~\ref{sec:algorithms}, we observe that the performance of treemaps depends on the characteristics of the datasets used. We identify four potential representative features that characterize time-dependent hierarchical datasets and classify all datasets used in our experiments accordingly. We experimentally test the validity of this classification on 2405 datasets, and analyze the relative performance of 14 state-of-the-art rectangular treemapping algorithms across varying features. Generally we conclude that our proposed features do indeed have predictive value, both with respect to visual quality and stability. We also observe that algorithms that are designed to be stable tend to in fact be more stable across features.

\noindent
\textbf{(3)} We perform a quantitative evaluation of 14 rectangular treemapping algorithms on more than 2000 datasets. We visually summarize our results with respect to both visual quality and stability to aid users in making an informed choice among treemaps. All datasets, metrics, and algorithms are openly available~\cite{URLTreemaps}. Section~\ref{sec:exploration} reports on our experimental results, we conclude in Section~\ref{sec:discussion}.

\section{Rectangular Treemaps}
\label{sec:algorithms}
We next discuss the most well-known rectangular treemapping algorithms. For a fair comparison during our experiments, we require that treemap rectangles have exactly the correct areas and partition the input rectangle. Algorithms that do not satisfy these requirements are not included in our evaluation. 
 Recall that the input for a rectangular treemap is a rectangle $R$ and a set of non-negative values $a_1, \ldots, a_n$ together with a hierarchy on these values (represented by a tree). The children of a node in this hierarchy are given in a particular order in the input. We distinguish two classes of treemaps, which either do or do not use this order. For time-dependent data we also distinguish between state-aware and stateless treemaps. Contrary to stateless treemaps, state-aware treemaps do not compute the treemap separately at a time step, but (can) use the layout of the previous time step to compute a new layout. Most treemaps are stateless; we discuss the state-aware algorithms separately. 

\smallskip\noindent
\textbf{Unordered treemaps} do not (need to) adhere to the input nodes' order when computing the layout. Typically, input weights are sorted to help the algorithm achieve good visual quality. Unordered treemaps in our evaluation include Squarified treemaps (SQR)~\cite{sqr} and Approximation treemaps (APP)~\cite{nagamochi2007approximation}. APP comes with a guaranteed upper bound on the worst-case aspect ratio, while SQR often achieves near-optimal aspect ratios in practice.
The visual quality of unordered treemaps is relatively unaffected by high \emph{weight variance}, as reordering weights allows the layout to group similar-size rectangles in the treemap, typically leading to better aspect ratios. Yet, the sorted order of the weights may change rapidly over time, especially if the \emph{weights change} much over time or if the \emph{weight variance} is low. This can negatively affect the stability of the treemaps. 

\smallskip\noindent
\textbf{Ordered treemaps} are required to adhere to the order of nodes as given in the input, which roughly ensures that rectangles close to each other in the input are close to each other in the resulting treemap. This typically improves the stability of treemaps, but may worsen visual quality. We include nine ordered treemaps in our evaluation. The first ordered treemaps~\cite{ordered} include the Pivot-by-Middle (PBM), Pivot-by-Size (PBZ), and Pivot-By-Split (PBS) algorithms. Similar algorithms are the Strip algorithm (STR)~\cite{bederson02} and the Split algorithm (SPL)~\cite{engdahl}. Other algorithms, like the Spiral algorithm (SPI)~\cite{spiral}, and the Hilbert (HIL) and Moore (MOO) algorithms~\cite{hilbert_moore}, lay out rectangles following a space-filling curve. Finally, the very first treemap algorithm (Slice-and-Dice (SND)) by Shneiderman~\cite{shneiderman92} can also be considered an ordered treemap. While not designed to be ordered, the resulting (combinatorial) layout depends only on the hierarchy and not on weights. In fact, SND uniquely uses the depth in the hierarchy to compute the layout (slicing vertically on even depth and horizontally on odd depth), rather than simply applying the same algorithm recursively. Hence, SND's visual quality strongly depends on the \emph{number of levels} in the input hierarchy. 
Typically, laying out large rectangles near small rectangles leads to poor aspect ratios. Hence, the visual quality of ordered treemaps is negatively affected by high \emph{weight variance}. However, ordered treemaps are relatively stable over time compared to ordered treemaps, as order is maintained. Finally, \emph{insertions and deletions} may affect the visual quality and stability of ordered treemaps to varying degrees, depending on how they are handled exactly.

\smallskip\noindent
\textbf{State-aware treemaps}
can use the layout of the previous time step to compute a new layout and so can largely control their stability. 
The treemap for the first time step is typically an existing unordered treemap. 
The first state-aware treemap was introduced by Sondag~\etal~\cite{sondag17}. Their Local Moves algorithm (LM) is initialized with APP, and allows only a small number of local modifications to the (combinatorial) layout between time steps. They also show how to update areas between time steps without significantly changing the layout (layouts remain ``order-equivalent''). In our evaluation we include the Local Moves algorithm with 4 local moves between time steps (LM4), and without local moves (LM0). A similar algorithm is the Git algorithm (GIT) by Vernier~\etal~\cite{vernier18git}, which is initialized with SQR, and does not allow any changes to the (combinatorial) layout between time steps. Both state-aware treemaps also support insertions and deletions, updating the layouts locally where necessary (for insertions, the position in the layout can be chosen to maximize visual quality).
By design, the stability of state-aware treemaps is relatively unaffected by frequent \emph{weight changes} over time. Also, the visual quality of the initial treemaps should be relatively high. However, since the layouts cannot change much over time, the visual quality of state-aware treemaps will decrease over time if weights change significantly. Frequent \emph{insertions and deletions} may also cause treemaps with poor visual quality, as treemaps are not recomputed as a whole. However, many insertions can help to correct rectangles with bad aspect ratio caused by weight changes over time. This is especially helpful for state-aware algorithms that do not allow any changes to the layout, like LM0 and GIT. Note that SND has a fixed layout if the input hierarchy does not change and is hence very stable, but it does not explicitly use the previous state.

Finally, note that the \emph{number of levels} in the input hierarchy can have a strong effect on all classes of algorithms. In general, more levels imply less freedom in the layout strategy. As a result, unordered treemaps become more similar to ordered treemaps. Overall, the visual quality tends to decrease and the stability tends to increase. 

\section{Metrics}
\label{sec:metrics}
There are two important criteria to evaluate treemaps: \emph{visual quality} and \emph{stability}. We discuss well-established metrics for both below. We also introduce a new method to measure the stability of time-dependent treemaps which captures inherent data changes. We measure metrics for each leaf rectangle separately and then aggregate these values for each algorithm and dataset~\cite{URLTreemaps}. Note that we do not compute metrics for non-leaf nodes. 

\subsection{Visual quality}
\label{sec:aspectratio}
The weight information in a treemap is conveyed by the areas of its rectangles. Since areas of rectangles closer to squares are visually easier to estimate than areas of elongated rectangles, visual treemap quality is commonly measured by the aspect ratio of its rectangles. Although it has been proposed that the ratio should be close to the golden ratio~\cite{lu2017golden} we aim for the uniformly accepted goal of making rectangles as square as possible. For a rectangle $R_i$ of width $w(R_i)$ and height $h(R_i)$, we define the aspect ratio $\rho(R_i)$ as
\begin{equation}
\rho(R_i) = \min(w(R_i), h(R_i)) / \max(w(R_i), h(R_i)).
\label{eqn:rho}
\end{equation}
Observe that this definition is the inverse of the usual definition for aspect ratio. Its values range from 0 to 1, where values of $\rho$ close to $0$ are considered ``bad'' and values close to $1$ are considered ``good''. The bounded range allows for easy aggregation. Note that, compared to the usual definition of $1/\rho$, rectangles with larger aspect ratios have a smaller influence on the aggregated score.

\begin{figure*}[b]
\centering
        \includegraphics[width=0.7\textwidth]{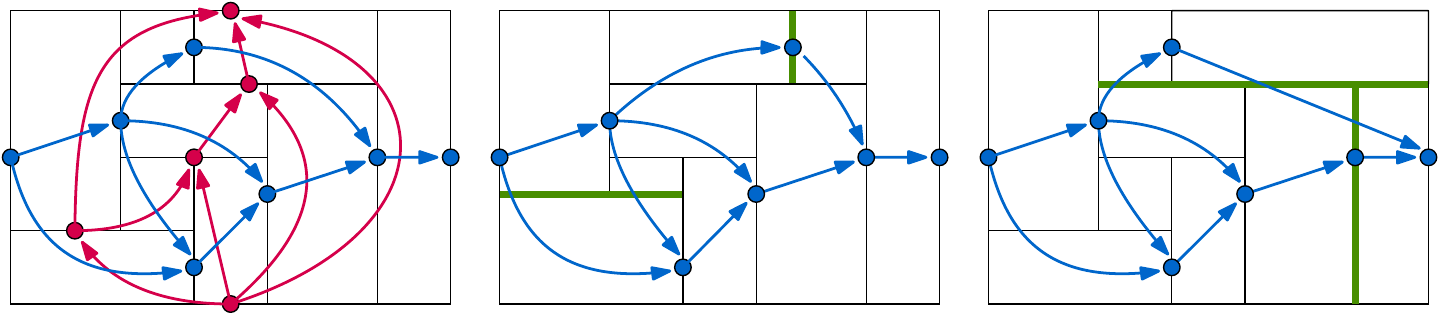}
    \caption{Left: Partial orders of the maximal segments. Middle: a layout order-equivalent to the left figure, changed maximal segments highlighted in green. Right: a layout not order equivalent to the other two figures. Red/blue arrows: relations between maximal segments.}
     \label{fig:partialOrder}
\end{figure*}

\subsection{Stability}\label{sec:stablity}
Evaluating the stability of a treemap is more involved than evaluating visual quality. Consider treemaps at two consecutive time steps $T(t)$ and $T(t+1)$. Since stability does not \emph{explicitly} depend on the value of $t$, we denote the former and the new treemap by $T$ and $T'$ respectively, to simplify notation. We also denote the rectangle areas in $T$ and $T'$ by $\{a_1, \ldots, a_n\}$ and $\{a'_1, \ldots, a'_n\}$, respectively. For a stable treemapping algorithm, the (visual) difference between $T$ and $T'$ should roughly correspond to the difference between $\{a_1, \ldots, a_n\}$ and $\{a'_1, \ldots, a'_n\}$. Note that the combination of large changes in data values and small changes in the layouts is unlikely since rectangle areas in treemaps must exactly match the data values. Hence, we want actually to measure \emph{instability}, that is, large layout changes that are not caused by large data changes.

Most existing treemap stability metrics consider only the visual change in the treemap's layout $d(T, T')$, usually computed by evaluating the change $\delta(R_i, R'_i)$ for each rectangle separately and aggregating it over all rectangles. Shneiderman and Wattenberg~\cite{ordered} define $\delta$ as the Euclidean distance between the vectors $(x(R_i), y(R_i), w(R_i), h(R_i))$ and $(x(R'_i), y(R'_i), w(R'_i), h(R'_i))$, where $x$, $y$, $w$, and $h$ are the coordinates of the top-left corner, width, and height of a rectangle, respectively. They then define $d$ as the average over all rectangles. Hahn~\etal~\cite{hahn10,hahn2015comparing} simplify this metric by defining $\delta$ as the distance moved by the centroid of a rectangle, again defining $d$ as the average. Tak and Cockburn~\cite{hilbert_moore} use the same $\delta$ as~\cite{ordered}, but define $d$ as the variance over all values computed by $\delta$. They also propose a drift metric, which measures how much a rectangle moves away from its average position over a long period. Recently, Scheibel~\etal~\cite{Scheibel2018} introduced two new layout change metrics: The \emph{average aspect ratio change} defines $\delta$ as the relative change between the aspect ratios of $R_i$ and $R'_i$, and defines $d$ as the average. The \emph{relative parent change} defines $\delta$ as the relative change of the distance between the center of a rectangle and the center of its parent, again defining $d$ as the average. Chen~\etal~\cite{Chen2017} propose a metric to quantify the ability of users to track time-dependent data in treemaps, which is closely related to the drift metric~\cite{hilbert_moore}. A different approach measures layout change using pairs of rectangles. Hahn~\etal~\cite{Hahn2017} introduce the \emph{relative direction change}, which, for every pair of rectangles $R_i$ and $R_j$, measures how much the angle from the center of $R_i$ to the center of $R_j$ changes. Finally, Sondag~\etal~\cite{sondag17} proposed the \emph{relative position change}, which, for every rectangle pair $(R_i, R_j)$, measures how much the relative position of $R_i$ with respect to $R_j$ changes. The distance $d$ is then defined as the average over all rectangle pairs. 

Summarizing the above, we distinguish two types of layout change metrics: (1) \emph{absolute} metrics measure how much individual rectangles move/change, and (2) \emph{relative} metrics measure how much positions of pairs of rectangles change relative to each other. For our experiments, we use both an absolute and a relative metric. In particular, as an absolute metric, we use the \emph{corner-travel distance}, which is a well-known metric used in computer vision to quantify change between two shapes using feature points~\cite{tuytelaars07,szeliski10}. In the vision community, it was established already many years ago~\cite{shi1994good,biederman87} that corners are a perceptually useful feature to identify and track.
Besides this perceptual validation, the corner-travel metric lies also within a small bounded factor of the original metric introduced by Shneiderman and Wattenberg~\cite{ordered}. Specifically, let $w(R)$ and $h(R)$ be the width and height of an input rectangle $R$, respectively. Let $p_i$, $q_i$, $r_i$, and $s_i$ ($p'_i$, $q'_i$, $r'_i$, and $s'_i$) be the positions of the corners of a rectangle $R_i$ ($R'_i$). We define the corner-travel (CT) distance for a rectangle as
\begin{equation}
\delta_{\text{CT}}(R_i, R'_i) = \frac{\|p_i - p'_i\| + \|q_i - q'_i\| + \|r_i - r'_i\| + \|s_i - s'_i\|}{4 \sqrt{w(R)^2 + h(R)^2}}.
\label{eqn:delta_ct}
\end{equation}
where $\|x\|_1$ denotes the $\ell_1$ norm. Simply put, $\delta_{\text{CT}}$ is the corner-to-corner correspondence distance between $R_i$ and $R'_i$. Note that $0 \leq \delta_{\text{CT}}(R_i, R'_i) \leq 1$, since a rectangle corner can travel by at most the length of the diagonal of $R$.

As a relative metric, we use the relative position change~\cite{sondag17}. We established experimentally that the corner-travel and the relative position change metric correlate clearly on more than 2000 data sets. Hence in Section~\ref{sec:exploration} we report only on experiments using the corner-travel distance. All other data can be found here~\cite{URLTreemaps}.

\mypar{Data change}
The stability metrics discussed above do not take data change into account. If data changes by a large amount, then the layouts should be allowed to change significantly without considering this to be instability. To add data change to a stability metric, one can consider the difference or ratio between the layout change and the data change~\cite{vernier18software,vernier18git}.
However, there are two problems: (1) we need a way to measure data change, and (2) the metric spaces for data and layouts need to be comparable. For example, data change can be measured in terms of changes of rectangle \emph{areas} (since these correspond to the data). However, layout changes such as the corner-travel distance measure \emph{lengths}, not areas. Areas and lengths are not directly comparable, and thus their ratios or differences may not be meaningful. 
Although such metrics could be made comparable by suitable normalization, such adaptations are necessarily metric-specific.

\mypar{Baseline treemap} We overcome the above issues with a new method that captures data change \emph{in the layout space}. For this, we define a \emph{baseline} treemap $T^*$ with respect to $T$ and $T'$. The layout of $T^*$ (that is, the combinatorial structure of the rectangular subdivision which constitutes $T^*$) is based on the layout of $T$. However, the areas of the rectangles in $T^*$ are the areas $\{a'_1, \ldots, a'_n\}$ of $T'$. The idea is that $T^*$ minimizes the layout distance to $T$ among all treemaps with the areas of $T'$. Put differently: $T^*$ represents the minimum amount of change that any time-dependent treemap must incur when moving from $T$ and its associated area values $\{a_i\}$ to the next treemap $T'$ and its area values $\{a'_i\}$. As a result, $d(T, T^*)$ is a good metric for data change in the layout space.

We construct $T^*$ for each tested algorithm and each time step using a hill-climbing algorithm, which was proven to converge in~\cite{eppstein2009area}. For a rectangular layout (treemap) $T$, a \emph{maximal segment} is a maximal contiguous horizontal or vertical line segment contained in the union of the borders of all rectangles in $T$ (for example, the green segments in Figure~\ref{fig:partialOrder}). Put simply, a horizontal maximal segment (which is not part of the input rectangle $R$) always has endpoints on the interior of two vertical segments and vice versa. For two horizontal maximal segments $s_1$ and $s_2$, we say that $s_1 < s_2$ if there is a rectangle in $T$ whose bottom side coincides with $s_1$ and whose top side coincides with $s_2$. This defines a partial order on horizontal maximal segments. We define a partial order on vertical maximal segments analogously (Figure~\ref{fig:partialOrder}). We say that $T$ is \emph{order-equivalent} to $T^*$ if the corresponding partial orders on maximal segments are isomorphic. For every possible set of areas, there exists an order-equivalent treemap to $T$ that correctly represents those areas~\cite{eppstein2009area}. In particular, we can initially define $T^*$ as the treemap order-equivalent to $T$ (computed with any of the tested algorithms) with the areas $\{a'_1, \ldots, a'_n\}$ of $T'$.

\begin{figure}[t]
\centering
\includegraphics[width=0.4\textwidth]{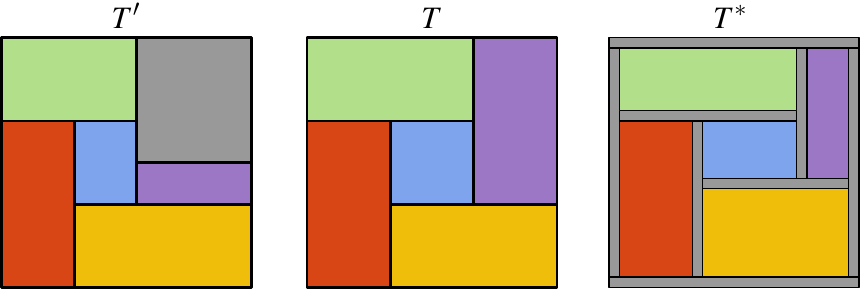}
    \caption{Treemaps $T'$ (with gray rectangle inserted), $T$, and $T^*$ (with gray area spread over maximal segments).}
     \label{fig:baseline}
\end{figure}

If rectangles are inserted or deleted, the baseline treemap cannot be order-equivalent to $T$, so we handle insertions and deletions separately. Dealing with deletions is easy: we simply let the areas go to zero. For insertions, we must be more careful. Indeed, while we consider only rectangles present in both $T$ and $T'$ when measuring stability ($R_i$ and $R_i'$ in Equation~\ref{eqn:delta_ct}), inserted rectangles can strongly impact the positions of rectangles in $T^*$. We observe that the baseline treemap does not need to be a proper treemap: it only needs to capture how much rectangles must minimally move to update to the new data. To minimize the movement of the rectangles due to insertions (and hence be as stable as possible), we distribute the cumulative area of the inserted rectangles over the ``walls'' (borders) of treemap $T$ as evenly as possible. To do so, we replace every maximal segment $e$ in $T$ by a rectangle, and assign each such rectangle a portion of the inserted area corresponding to the length of $e$ (Figure~\ref{fig:baseline}). Hence all walls become equally thick and the original rectangles of $T$ need to move as little as possible to yield $T^*$.

\begin{figure}[b]
    \centering
    \includegraphics{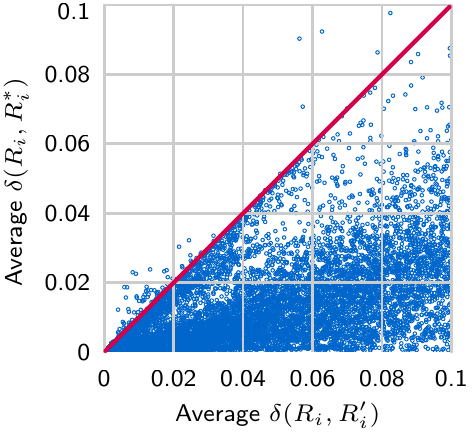}
    \caption{Scatter plot of the average layout change between $T$ and $T'$ or $T^*$ for a random 25\% sample of all algorithms and datasets.}
    \label{fig:CTDvsBase}
\end{figure}

The baseline treemap $T^*$ does not necessarily minimize the movement of every rectangle. However, the layout change between $T$ and $T^*$ is still a very good estimate for the minimum necessary layout change between $T$ and $T'$, and thus a good measure for data change (see Figure~\ref{fig:CTDvsBase}: nearly all points lie on or below the diagonal). 

%

\mypar{Stability metric}
We can now define a stability metric that takes data change into account. Consider a rectangle $R_i$ and the corresponding rectangles $R'_i$ and $R^*_i$ in $T'$ and $T^*$, respectively, and let $\delta$ be the layout change function for single rectangles. Two natural choices for spatial stability are the difference or ratio between $\delta(R_i, R'_i)$ and $\delta(R_i, R^*_i)$. Our experiments showed that the difference is typically more informative, that is, it exhibits clearer, more pronounced patterns, than the ratio. Hence, we define the stability of a single rectangle 
as
\begin{align}
\sigma(R_i) &= \max(0, \delta(R_i, R'_i) - \delta(R_i, R^*_i)) \label{eqn:sigma_ct}
\end{align}
Note that $\sigma(R_i) = 0$ if $\delta(R_i, R'_i) \leq \delta(R_i, R_i^*)$, which is possible. Indeed, a value of $0$ for $\sigma(R_i)$ represents ``very stable'', and $R_i^*$ is considered to be (roughly) as stable as possible.

\mypar{Limitations} The stability metrics we use focus only on consecutive time steps. The stability of time-varying treemaps could conceivably be influenced by effects that span multiple time steps, which our metrics do not capture directly. However, we believe that the most salient events influencing stability occur between consecutive time steps and hence we focus on this scenario.

\section{Data}
\label{sec:data}
The visual quality and/or stability of treemaps clearly depends on the datasets used. Simply measuring the average performance over a (large) collection of datasets does not reveal such information. We aim to provide sufficient insight so that both practitioners and researchers can make informed choices about which algorithm to use for their data. For this, we study the performance of treemaps as a function of the characteristics of the input data. We classify the datasets into data classes along with explicit features and evaluate the metrics of different treemapping algorithms for each class. 

\subsection{Data features}
\label{sec:dataspace}
Our methodology is inspired by the framework proposed by Smith-Miles~\etal~\cite{smith2014towards} to objectively measure the performance of algorithms across datasets. For each dataset, we compute a number of features that (hopefully) capture the characteristics influencing the relative performance of treemapping algorithms. As a result, every dataset is represented by a point in a low-dimensional \emph{feature space} $\mathcal{F}$. Similar feature-based approaches are also used to measure the relative performance of dimensionality-reduction methods~\cite{espadoto19} or in machine learning~\cite{bishop06}. Based on the discussion of treemapping algorithms in Section~\ref{sec:algorithms}, we identify the following four features: \textbf{1.} Levels of hierarchy, \textbf{2.} Variance of node weights, \textbf{3.} Speed of weight change, and \textbf{4.} Insertions and deletions.

Obviously, other features can be used to characterize (time-dependent) trees, such as the total node count; the min, max, and average node degrees; and the (im)balance of the tree structure~\cite{boorman73,kuhner14}. However, for a practical and meaningful analysis, we must keep the number of features low. We hence limit the analysis to the four features discussed above. 


\subsection{Data classes}
\label{sec:datasampling}
Using the feature space $\mathcal{F}$, we partition all datasets into classes. For each feature we define a small number of subclasses based on only that feature. The data class of a dataset is then defined as the combination of the subclasses for each feature. We determined the value-ranges defining the subclasses by analyzing the distribution of feature values over our 2405 real-world tree datasets.  


\mypar{Levels of hierarchy (3 subclasses)} We use three ranges for classification: 1 level (1L), 2 or 3 levels (2/3L), and more than 3 levels (4+L). Most hierarchical datasets we have analyzed have 2 or 3 levels. This number of levels is quite common for datasets that are visualized via treemaps, since they frequently concern geo-spatial subdivisions such as countries, continents, and their subregions, grouped by a classification scheme, such as the World Bank regional classification. Furthermore, visually understanding the node nesting in deeper treemaps becomes difficult~\cite{vliegen,sqr}.
A special case are datasets with only 1 level, that is, sets of weight values. Such datasets are also often visualized by treemaps, as these are more space-filling than alternatives such as bar charts~\cite{vliegen}. These datasets are challenging for treemaps that implicitly use the depth of the hierarchy. Finally, we consider datasets with more than 3 levels, which correspond to deep hierarchies such as, for example, file systems or software architectures~\cite{hahn10,Hahn2017,vernier18software}.

\mypar{Variance of node weights (2 subclasses)} We distinguish between low variance (LWV) and high variance (HWV). To ensure that the total number of tree nodes does not strongly influence our classification, we use the coefficient of variation $\sigma/\mu$ to determine the subclass. The standard deviation $\sigma$ and the mean $\mu$ are computed over all leaf weights over all time steps. We say that there is low variance if $\sigma/\mu \leq 1$ and high variance if $\sigma/\mu > 1$, respectively.

\mypar{Speed of weight change (3 subclasses)} We distinguish between low weight change (LWC), regular weight change (RWC), and spiky weight change (SWC). The weight change of a single rectangle is measured by the absolute difference in the relative area (with respect to the input rectangle R) between time steps. The weight change of a treemap between two time steps is defined as the sum of weight changes of all rectangles. To determine the subclass of a dataset, we use the distribution of weight changes between time steps over all time steps in the dataset, specifically the mean $\mu$ and the standard deviation $\sigma$. Datasets with low weight change have $\mu < 5\%$ and $\sigma < 5\%$. Datasets with a larger mean ($5\% \leq \mu < 20\%$) and a relatively small coefficient of variation ($\sigma/\mu \leq 1$) are classified as having regular weight change. The weights of these datasets steadily change over time, without any extreme changes. Remaining datasets are classified as having spiky weight change. In those datasets weights change drastically ($\mu > 20\%$), or there is large variation ($\sigma/\mu > 1$) along with substantial changes ($\mu > 5\%$ or $\sigma > 5\%$). 


\mypar{Insertions and deletions (3 subclasses)} We distinguish between low insertions and deletions (LID), regular insertions and deletions (RID), and spiky insertions and deletions (SID). We measure the impact of insertions and deletions between two time steps $t$ and $t+1$ as the cardinality of the symmetric difference between the two sets of rectangles with non-zero weights at $t$ and $t+1$, divided by the number of rectangles with non-zero weights at $t$. We again classify the datasets based on the distribution ($\mu$ and $\sigma$) of impact values over all time steps. Same as for the speed of weight change, LID is defined by $\mu < 5\%$ and $\sigma < 5\%$, RID is defined by $\mu < 20\%$ and $\sigma/\mu \leq 1$, and the remaining datasets are in SID.


\medskip
\noindent
The full classification results in $3 \times 2 \times 3 \times 3 = 54$ data classes. In Section~\ref{sec:exploration}, we evaluate how the performance of treemapping algorithms depends on the classes, that is, if the classification is sensible. 


\subsection{Datasets}
\label{sec:datasets}
\begin{figure}[t]
    \centering
    \includegraphics{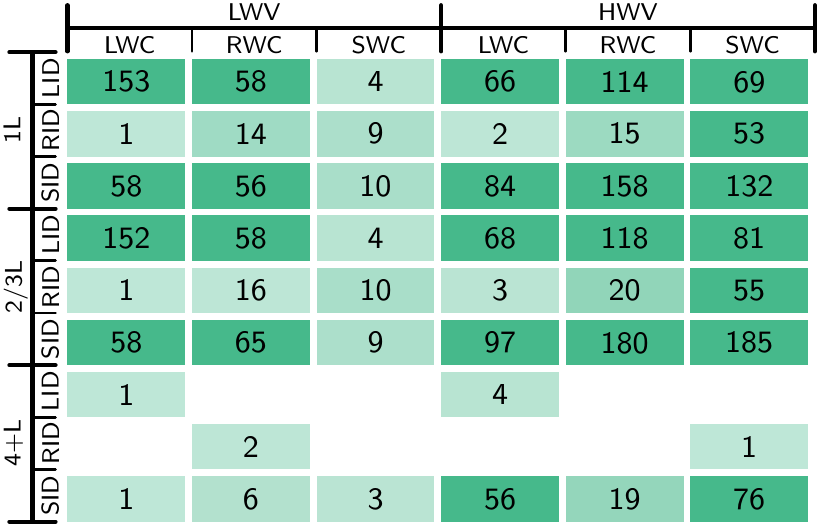}
    \caption{Distribution of datasets over classes.}
    \label{fig:datasets_summary}
\end{figure}

We collected a total of 2405 time-dependent hierarchical datasets from a variety of sources, detailed below. We found at least one dataset for 46 (out of 54) instances of our proposed data classes. See Figure~\ref{fig:datasets_summary} for the distribution of datasets over classes: clearly not all classes arise with equal frequency in our data sources.

\begin{description}
\item[Worldbank~\cite{URLWorldbank}:] World development indicators such as agriculture, rural and urban development, education, trade and health.
\item[GitHub~\cite{URLGit}:] Hierarchies of folders, files, and classes, weighted by the number of code lines, extracted from all revisions of several popular GitHub repositories using Scitools~\cite{URLscitools}.
\item[Movies:] Movies from MovieLens\cite{harper2016movielens} and \emph{TMDB}~\cite{URLmdb}. We constructed a time-dependent hierarchy using the group-rows-by-attribute-value partitioning method~\cite{tablelens,vliegen}. The hierarchy groups movies based on their genres, actors, release date, and keywords. Each leaf is a movie, whose weight corresponds to its rating over a given period of time.
\item[Custom:] We added several hand-picked datasets to cover classes that did not appear in the automatically mined datasets: \emph{Dutch Names}~\cite{urlmeertens} contains the frequency of popular baby names in the Netherlands per year; \emph{UN Comtrade Coffee}~\cite{URLComtrade} contains the amount of coffee each country imported per year; \emph{ATP} contains personal information, historical rankings, and match results from 1968 to 2018 for ATP tennis players~\cite{URLatp}; and \emph{Earthquakes} contains the time, location, depth and intensity of seismic phenomena provided by the USGS Earthquake Hazards Program~\cite{URLearth}.
\end{description}
Importantly, note that the above selection of dataset sources is \emph{orthogonal} to the description of the feature space $\mathcal{F}$. The former covers the \emph{origin} of data (which may cover application-specific aspects not captured by our feature space); the latter covers application-independent data aspects as captured by the data classes of $\mathcal{F}$.


\section{Experimental Results}
\label{sec:exploration}

We ran all 14 algorithms on all time steps of all 2405 datasets, generated the baselines for all these instances (Section~\ref{sec:metrics}), and recorded all layouts, that is, the positions of all rectangles $R_i(t)$ at all time steps $t$. Per dataset we aggregate our results for all metrics and algorithms by first by taking the mean over all rectangles in a single time step, and then by taking the mean again over all time steps. This is necessary since the number of rectangles may differ per time step. 





We focus on three specific questions: We explore the \emph{validity} of our data classification (Section~\ref{sec:validity}); we study the \emph{performance} of all algorithms with respect to visual quality and stability across varying data features (Section~\ref{sec:acrossfeatures}); and we compare the performance of all algorithms on each data \emph{class} separately (Section~\ref{sec:comparison}). We believe that the resulting visual summary will help researchers and practitioners 
choose a suitable treemapping algorithm for their data.

\subsection{Data classification analysis}\label{sec:validity}

We evaluate if the relative performance of treemapping algorithms is more consistent within a data class than for an arbitrary collection of datasets. To perform this analysis we need to establish how we can capture the consistency of relative performance for a collection of datasets, and how we can compare this consistency between multiple collections. 
We restrict our analysis to data classes that contain at least 50 datasets, for otherwise the observed consistency is not sufficiently reliable. For each such data class, we randomly sample 50 datasets to use in this analysis. We also randomly sample 50 datasets among all 2405 datasets (all classes) as a baseline for comparison. Note that all collections must have the same number of datasets in the analysis to ensure that the comparisons are fair.

Now consider a single collection of datasets. To measure the consistency of relative performance among different datasets in this collection, we cannot directly use the computed metrics for visual quality and stability, as these values may differ greatly between datasets. Alternatively, we could rank the algorithms per dataset, but then algorithms with very similar performance may imply a greater variance in relative performance than is the case. Instead, we define the \emph{relative performance} (separately for visual quality and stability) per dataset as follows. We compute both the best value (maximum for visual quality, minimum for stability) and the median value over all algorithms over this dataset. The \emph{relative performance score} for each algorithm on this dataset is then computed by linearly interpolating between these two values, where the best algorithm receives score $0$, and the median algorithm receives score $0.5$. The relative performance score is capped at $1$, to avoid outliers. The resulting scores are comparable between different datasets.

\begin{figure}[t]
    \centering
    \includegraphics{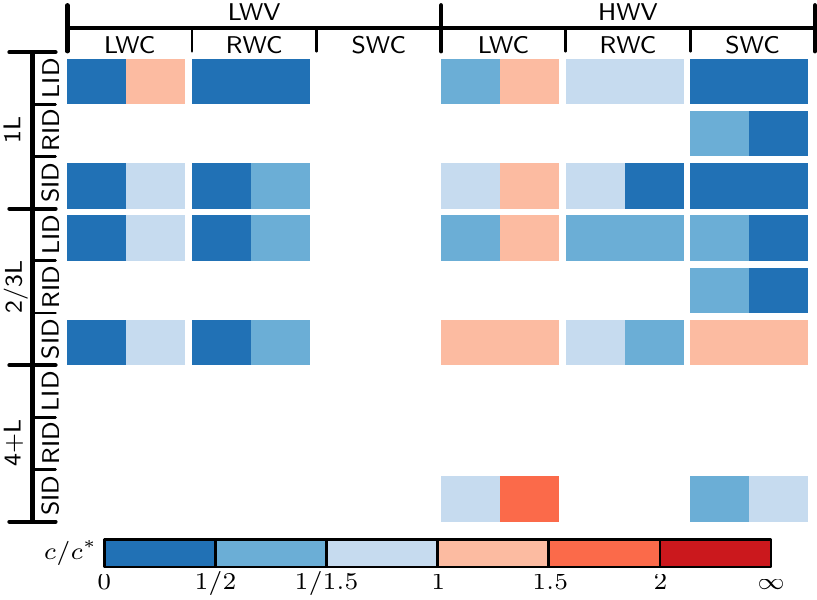}
    \caption{For each data class with at least 50 datasets, the ratio of the consistency score (visual quality on the left, stability on the right) between the data class and the baseline. }
    \label{fig:variancetable}
\end{figure}

We next analyze the consistency of relative performance within collections of datasets in two different ways. First, we use a quantitative approach: for each algorithm we compute the variance of the relative performance scores over all datasets in a collection, and we sum up these variances over all algorithms. This results in a \emph{consistency score} $c$ for a collection of datasets. Figure~\ref{fig:variancetable} displays the consistency scores (for visual quality and stability) of all data classes (with at least 50 datasets) compared to the consistency scores $c^*$ of the baseline collection (created by random sampling). A cell is colored blue (more consistent) if $c$ is smaller than $c^*$; a cell is colored red (less consistent) if $c$ is larger than $c^*$.

Nearly all data classes for visual quality and most data classes for stability are more consistent than the baseline. This indicates that our features are splitting the datasets into valid data classes where the relative performance of an algorithm is easier to predict than in the baseline.
However, the stability column for high weight variance and low weight change is less consistent than the baseline. As discussed in Section~\ref{sec:algorithms}, the stability of unordered treemaps becomes worse compared to ordered treemaps when the weight variance is low or the weight change is high, due to reordering of the input weights. As a result, the difference with respect to stability between ordered and unordered treemaps is less pronounced for these data classes; the relative performance is hence influenced more by accidental details of individual datasets and less by structural differences between the algorithms. Additionally there are two data classes where the visual quality is less consistent than the baseline. 
It is not clear to us at this point what the cause of these inconsistencies is; one possibility are hidden correlations in the data classes.

\begin{figure*}[t]
    \centering
    \includegraphics{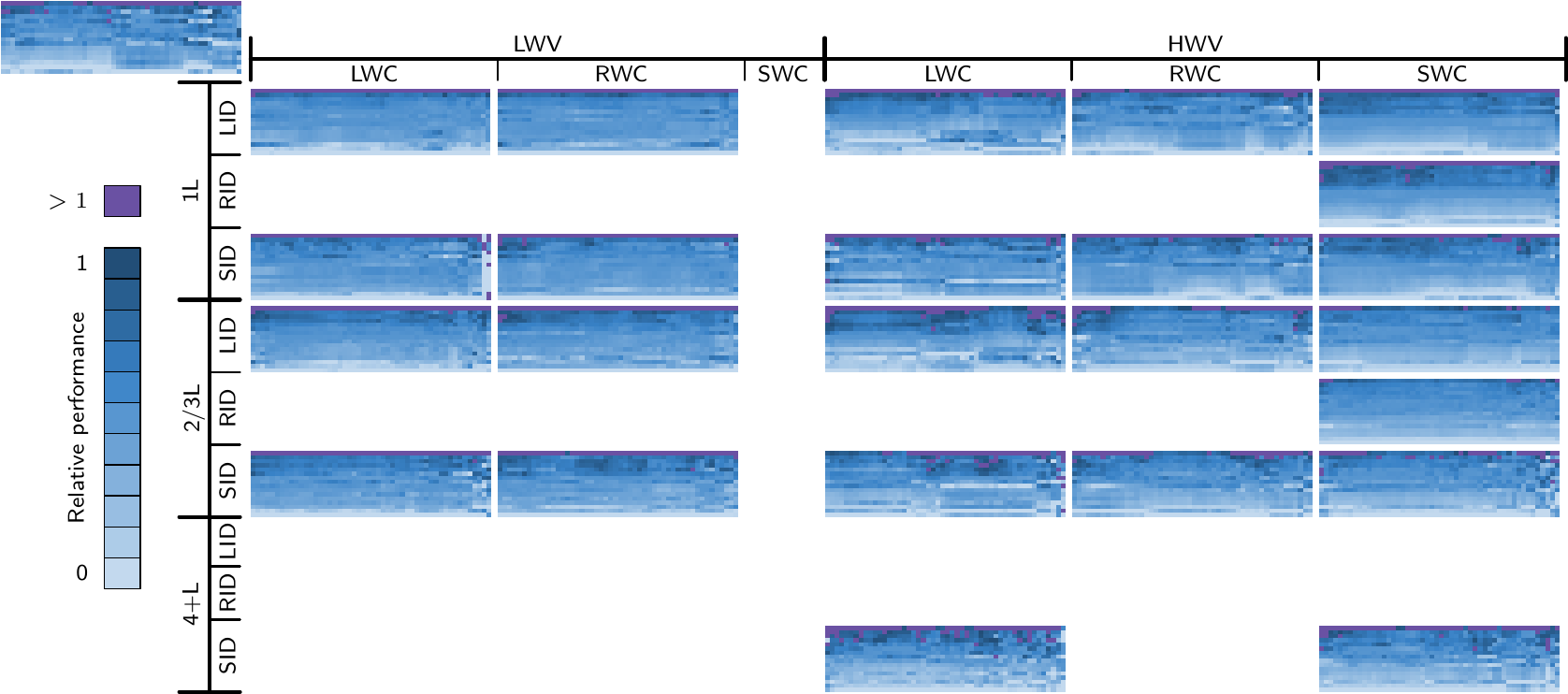}
    \caption{Visual quality: matrix plots for each data class with at least 50 datasets plus baseline (left top). In each matrix plot, rows correspond to algorithms, columns to datasets. The lighter the color, the better the relative performance, capped at 1 (purple).}
    \label{fig:meanrug}
\end{figure*}

\begin{figure*}[t]
    \centering
    \includegraphics{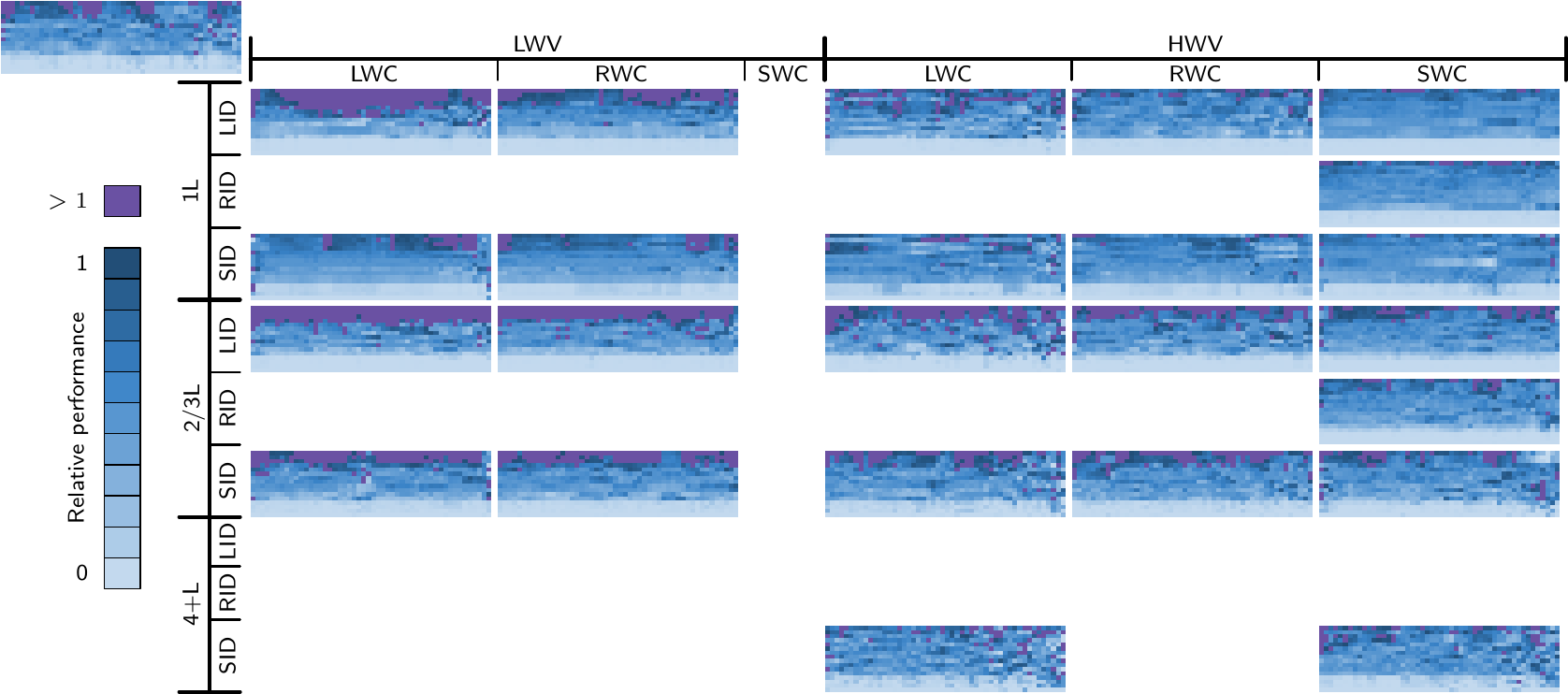}
    \caption{Stability: matrix plots each data class with at least 50 datasets plus baseline (left top). In each matrix plot, rows correspond to algorithms, columns to datasets. 
    The lighter the color, the better the relative performance, capped at 1 (purple).}
    \label{fig:ctdrug}
\end{figure*}

Second, we use a more qualitative approach to assess the consistency of relative performance. For each data class we create a matrix plot, that shows the relative performance scores of all algorithms for all datasets in the collection (see Figures~\ref{fig:meanrug} and~\ref{fig:ctdrug}). Each column in the matrix plot represents a dataset, and each row represents an algorithm. The color of every ``cell'' in the matrix plot indicates the relative performance score of an algorithm on a dataset, where lighter colors indicate better (lower) relative performance scores. Relative performance scores that were capped at $1$ are indicated with purple. To better enable the visual assessment of consistency among the different datasets in a collection, we order the datasets (columns) so that those with similar scores are next to each other as much as possible. Also, we order the algorithms (rows) so that the algorithms with better average score are lower in the matrix plot. In particular, the order of algorithms in the matrix plots for different data classes can be different. Figure~\ref{fig:meanrug} shows the matrix plots for visual quality, with the corresponding matrix plot for the baseline collection at the left-top. Figure~\ref{fig:ctdrug} shows the matrix plots for stability.


First consider the matrix plot for visual quality (Figure~\ref{fig:meanrug}). For the low weight variance subclass we indeed see that the matrix plots are much smoother than the baseline, which confirms the results in Figure~\ref{fig:variancetable}. 
%
We also observe an increasing number of irregularities when going from 1 level treemaps to 2/3 levels or 4+ levels, since more levels impose more restrictions on the layout and hence all algorithms perform more similarly. 

Consider now Figure~\ref{fig:ctdrug}. First of all, we notice that there is a set of four algorithms at the bottom of every matrix plot. These are the state-aware algorithms and SND. For nearly all datasets, regardless of the specific data class, these four algorithms are much more stable than any of the others. 
There is a large difference between the low weight variance and high weight variance subclasses. For low weight variance there is a set of algorithms that perform consistently much worse than the median (purple cells). These include the unordered treemaps which are particularly sensitive to changes in such data.

\subsection{Performance analysis across features}
\label{sec:acrossfeatures}
\begin{figure}[t]
    \centering
    \includegraphics[width=\linewidth]{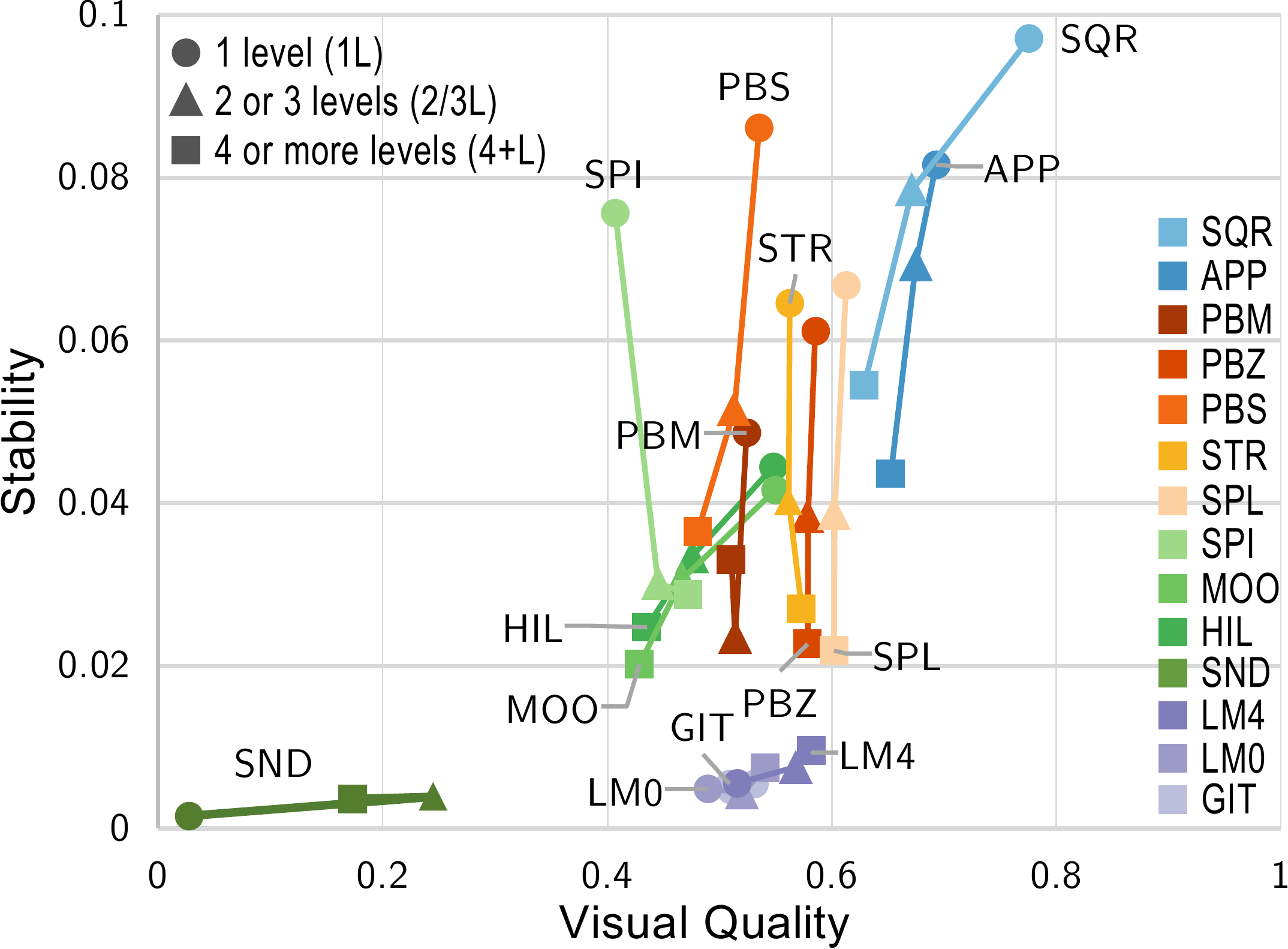}
    \caption{Visual quality \emph{vs} stability as function of the levels of hierarchy feature.}
    \label{fig:depth}
\end{figure}

\begin{figure}[b]
    \centering
    \includegraphics[width=\linewidth]{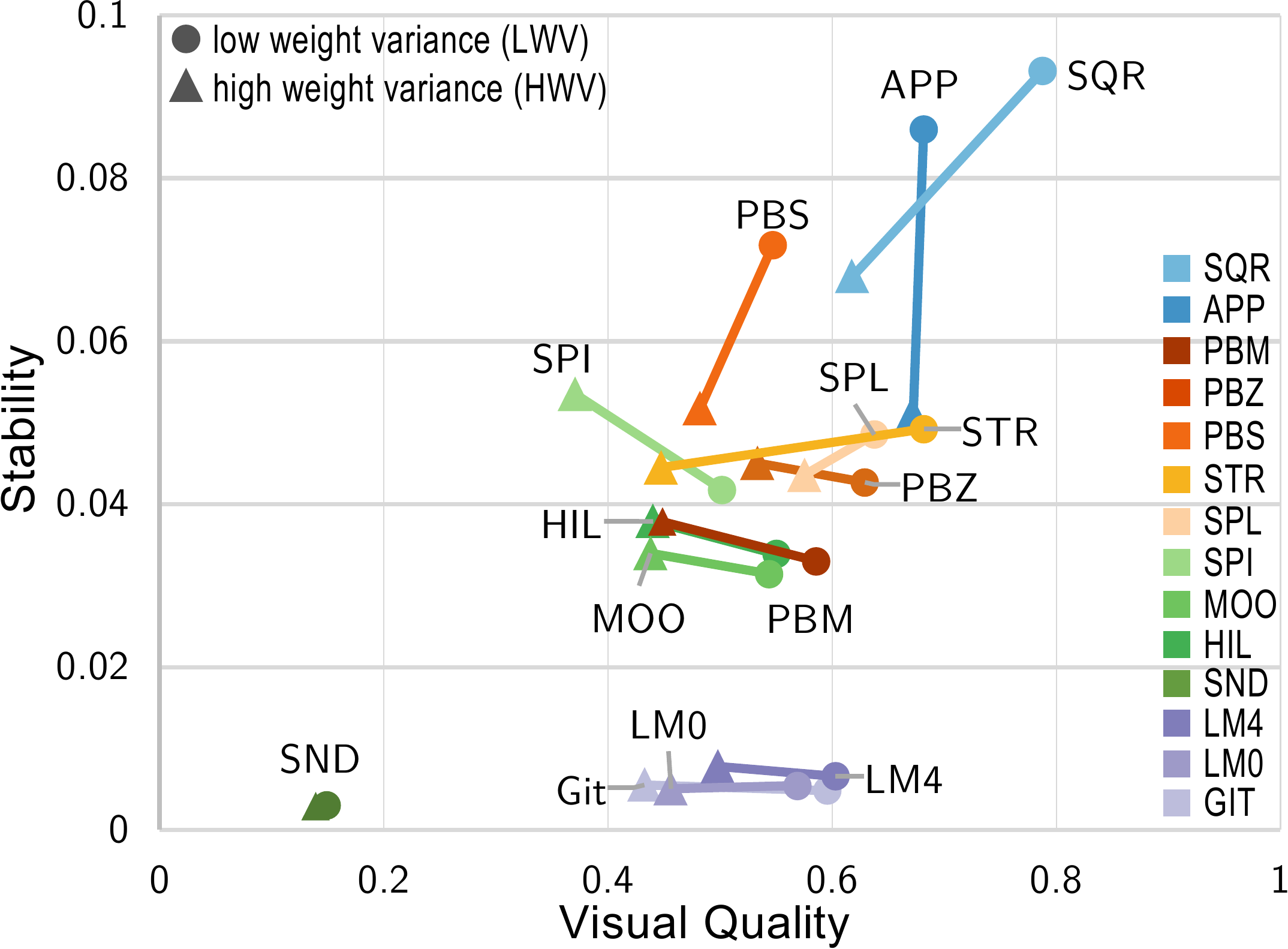}
    \caption{Visual quality \emph{vs} stability as function of the variance of node weights feature.}
    \label{fig:weightVariance}
\end{figure}

\begin{figure}[t]
    \centering
    \includegraphics[width=\linewidth]{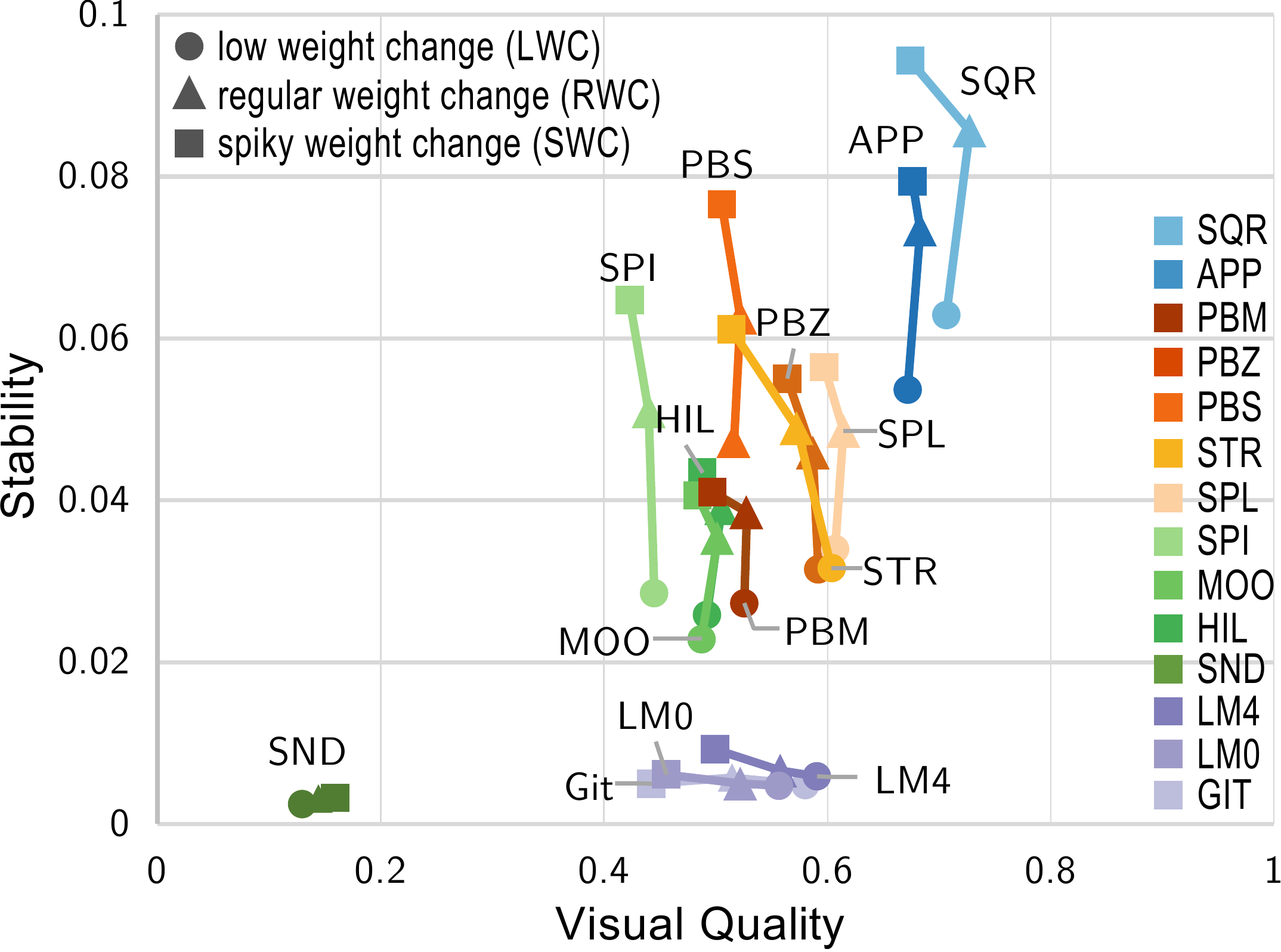}
    \caption{Visual quality \emph{vs} stability as function of the speed of weight change feature.}
    \label{fig:weightChange}
\end{figure}

\begin{figure}[t]
    \centering
    \includegraphics[width=\linewidth]{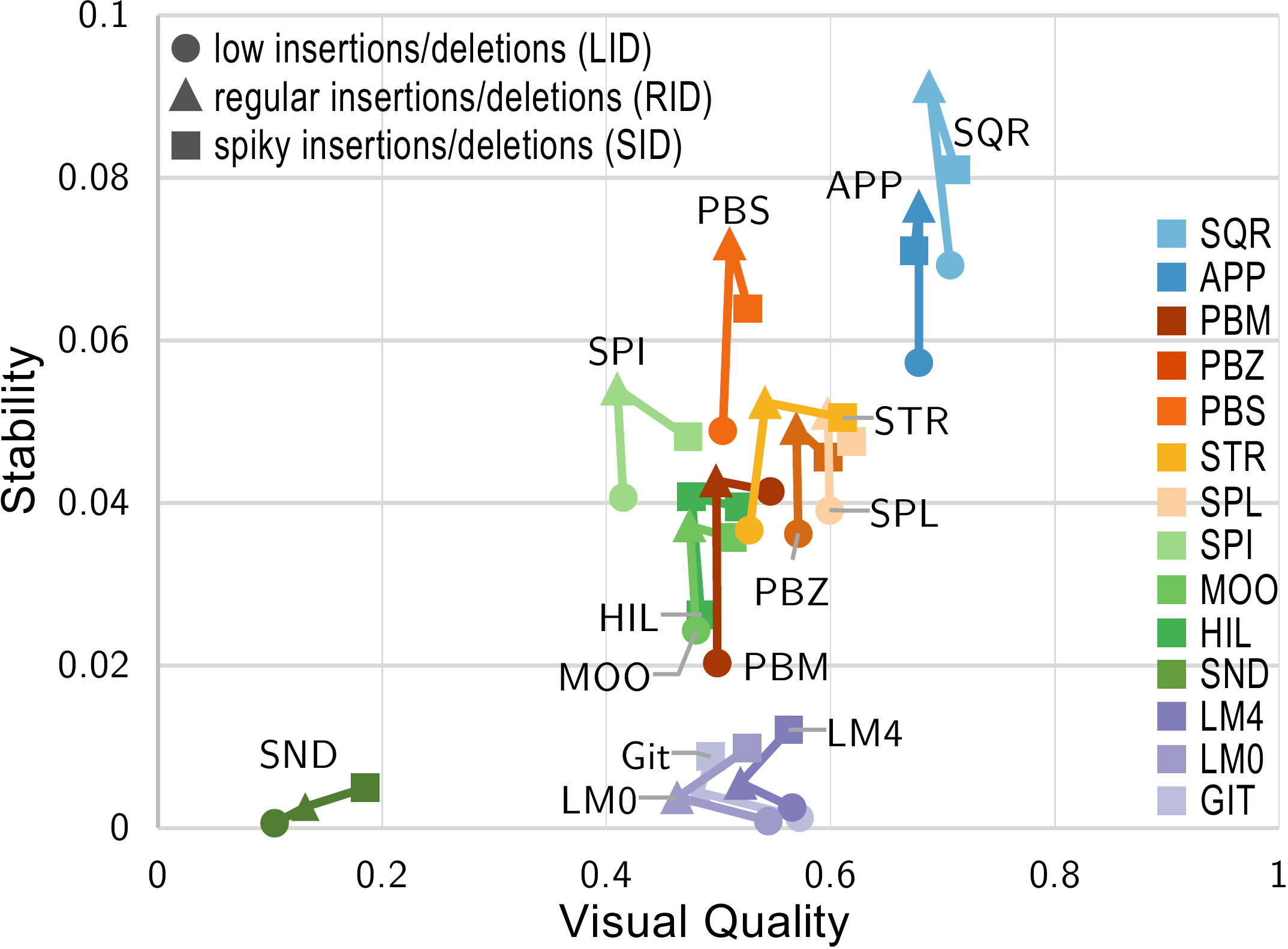}
    \caption{Visual quality \emph{vs} stability as function of the insertions and deletions feature.}
    \label{fig:insertionsDeletions}
\end{figure}

%
%
The analysis in Section~\ref{sec:validity} shows that our data classification is valid. We now study how visual quality and stability depend on the \emph{features} of the datasets. 
We aim to understand how sensitive a given algorithm is to variations in one or several features of its input data.
For each data class we calculate the average visual quality and stability. For each subclass of a feature we then take the average over all data classes that belong to it. This ensures that even though we have different numbers of datasets in each data class, they are all weighted equally. We show this data in Figures~\ref{fig:depth},\ref{fig:weightVariance},\ref{fig:weightChange},\ref{fig:insertionsDeletions}. Each point in each figure represents the score for one algorithm on one subclass of the feature, for example, low weight variance. We draw a polyline that connects the points of one algorithm and use glyphs to indicate the different subclasses. The different algorithms are indicated with categorical colors (see figure legends).

Recall that a low value for the stability metric indicates a \emph{stable} algorithm and that the visual quality metric (aspect ratio) is bounded between 0 and 1. In particular, note that visual quality ($\rho$) of 0.5 for a single rectangles indicates a 2-by-1 rectangle.  
A $\rho$ of 0.25 however is perceptually much worse than a $\rho$ of 0.5 in terms of area perception as can be inferred from Kong~\etal~\cite{Kong2010}, coming close to their "extreme aspect ratios" of 4.5.

\mypar{Levels of hierarchy} Figure~\ref{fig:depth} considers the levels of hierarchy feature, which has three values: 1L, 2/3L, and 4+L. 
From Figure~\ref{fig:depth}, we see that all algorithms, in particular the stateless ones, are more stable as the number of levels increase. In contrast to most other algorithms, the visual quality of state-aware algorithms (LM0, LM4, GIT) as well as SND increases with the number of levels. 
We also see that SQR and PBS have the longest polylines, that is, they are the most sensitive to the number of levels.  

\mypar{Variance of node weights} Figure~\ref{fig:weightVariance} considers the weight variance feature, which has two values: LWV and HWV. 
Increasing the weight variance decreases the visual quality for all algorithms, except for APP (and SND). Additionally we see that the unordered treemaps are indeed more sensitive to this feature in terms of stability compared to the other algorithms. These algorithms reorder the data based on the weight to determine their layout, and if the weight are close to each other this happen more often.

\mypar{Speed of weight change} Figure~\ref{fig:weightChange} considers the speed of weight change feature, which has 3 values: LWC, RWC, and SWC. The near-vertical polylines for the stateless algorithms show that visual quality seems to be largely unaffected by this feature. The stability however decreases quickly. Conversely, for the state-aware algorithms the polylines are mostly near-horizontal: the stability is largely unaffected, but the visual quality decreases. As the only state-aware algorithm that allows changes to the layout, LM4 makes an explicit tradeoff between stability and visual quality, which is visible by the slightly sloping line.

\mypar{Insertions and deletions} Finally, Fig.~\ref{fig:insertionsDeletions} considers the insertions and deletions feature, which has three values: LID, RID, and SID.
The plot shows a similar variation of visual quality and stability as seen for the speed of weight change feature (Fig.~\ref{fig:weightChange}). 
Yet, the polylines for the stateless algorithms now show a `kink' at the midpoint (RID, regular insertions/deletions). This tells us that these algorithms are most unstable for regular insertions/deletions, and stabler for linear and spiky insertions/deletions. Interestingly, the state-aware methods (LM0, LM4, GIT) show a similar kink but oriented differently. These methods hence achieve poorest visual quality for regular insertions/deletions and highest quality on the other two values of this feature.

\subsection{Comparison of data classes}
\label{sec:comparison}

\begin{figure*}[htbp!]
\centering
\includegraphics[height=0.95\textheight]{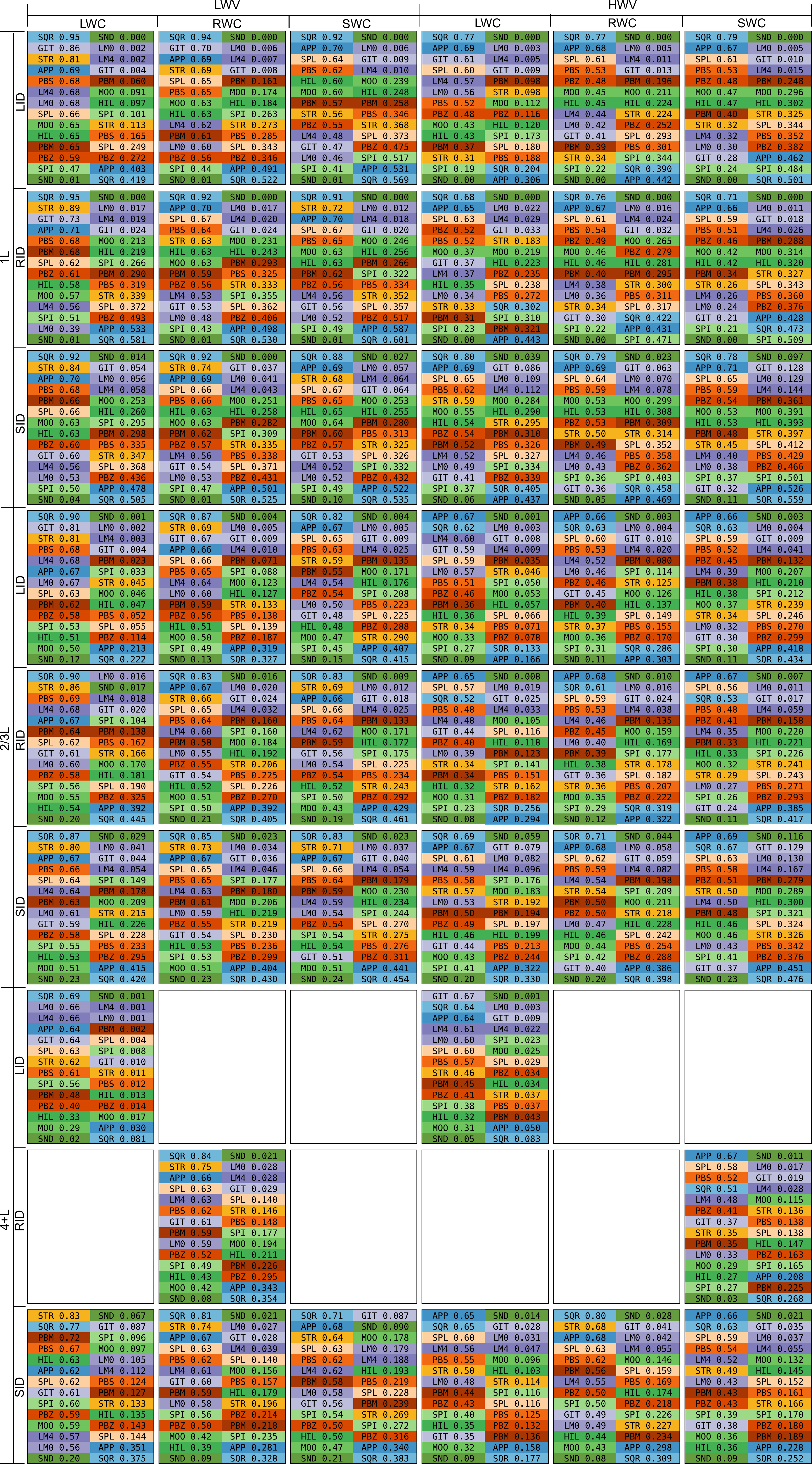}
\caption{Relative ranking of treemapping algorithms for all data classes. Each table cell shows algorithms in top-down decreasing order of average visual quality (left column) and average stability (right column).}
\label{fig:rankingtable}
\end{figure*}

We next compare the relative performance of all algorithms separately on all data classes. 
Figure~\ref{fig:rankingtable} supports this comparison as follows: it is structured as a matrix of tables, one per data class.
Each table shows the average visual quality (left column) and average stability (right column) of all algorithms for all datasets in the respective data class. The two columns are sorted separately to show the best-ranking algorithms at the top. Cells show the algorithm names and scores, and are categorically color-coded on the algorithm name, following the same color scheme as in Section~\ref{sec:acrossfeatures}. Empty cells indicate data classes for which we did not find datasets.
Figure~\ref{fig:rankingtable} can answer the following practical questions:
\begin{description}
\item[Which method is best for my data?] Given a family of datasets with known characteristics (feature values), we search for the corresponding cell and pick the top algorithm(s) in visual quality, stability, or a combination of both, depending on the application requirements. When doing this, we should examine the actual values, since several algorithms score quite close to each other.
\item[How is a given algorithm performing in general?] We scan the table following the color of the respective algorithm, and detect its rank (with respect to visual quality and/or stability) over all data classes. In this way we can find patterns and outliers in the data for this algorithm: for example, LM0 and LM4 are always near the top in stability, and GIT's performance on visual quality fluctuates widely depending on the data class.
\item[Which algorithms perform similarly?] We locate groups of neigh\-boring rows with the same color pattern in all tables. These indicate algorithms which score similarly regardless of data class.
\end{description}
%
In general there are a number of insights we can obtain from Figure~\ref{fig:rankingtable}. When we consider only the visual quality, we see that SQR is usually the best for low-weight variance data, but for high weight variance APP is just as often the best algorithm. If the dataset contains only 1 level, SQR performs better, but for the other depth subclasses it depends on the exact data class.
If only the stability is important, SND almost always scores best regardless of the data class, but likewise it consistently scores the poorest on visual quality. 
The state-aware algorithms all perform very well on stability. While LM0 is better in terms of stability than LM4, their exact order as well as their relative order to GIT varies depending on the data class.
When considering which algorithm is best for both stability and visual quality, there are no easy answers. There is no algorithm that performs best on both in any of the data classes and hence the answer depends on the desired trade-off and the data class in question.

\section{Discussion and Conclusion}
\label{sec:discussion}

We performed an extensive quantitative evaluation of rectangular treemaps for time-dependent data. To do so, we introduced a new methodology based on baseline treemaps to measure the stability of time-dependent treemaps. Baseline treemaps enable us to measure the change in the input data in a manner that is mathematically comparable to the measures for the layout change of the corresponding treemaps. Furthermore, we proposed a novel classification scheme for time-dependent data sets via a four-dimensional feature space (weight variance, weight change, tree depth, and the pattern of insertions and deletions). These four features naturally arose from a discussion on various types of state-of-the-art treemapping algorithms. Our experimental analysis shows that our proposed classification is valid in general and that most data classes are well suited to predict the performance of treemapping algorithms. For most data classes, our visual summary comparing all algorithms across all data classes and both metrics can hence serve as a reliable resource for researchers and practitioners. Last but not least, all datasets, metrics, and algorithms used in our evaluation are openly available~\cite{URLTreemaps}.

\mypar{Limitations and future work} Our experiments show that our identified features and the resulting feature space generally work well and result in a meaningful classification of datasets. However, there are whole sets of data classes for which we could not find sufficiently many (or even any) datasets. This is partially inherent in the classification and somewhat natural: data sets with low weight variance hardly ever exhibit spiky weight change behavior, so that particular column in our table is essentially empty. But among the 18 classes of treemaps with 4 or more levels we found a significant number of datasets only for two classes, which both are essentially populated by datasets stemming from software repositories. The question remains if there are other significant types of time-dependent hierarchical datasets which have four or more levels and which escaped our searches. As it is, the results for these two particular classes are representative for only a restricted type of data.

Our classification works well for visual quality, with the exception of two cases (2/3 level, spiky insertions and deletions, high weight variance, and low or spiky weight change). We have a large number and variety of datasets at our disposal for these two classes, but nevertheless, it is unclear to us what causes these inconsistencies in the performance of the tested algorithms. There might be a hidden correlation in these datasets and one or more additional features might be needed to separate these classes further.

While we do have a significant number of datasets at our disposal and hence can validate our claims with some certainty, we still might be observing some bias in our collection. We would hence like to construct, and evaluate on, synthetic datasets. Doing so is not trivial; creating datasets that avoid sampling biases and are representative of real-world datasets (for a suitable definition of ``real-world'') is a challenging (but important) question in its own right in information visualization in particular and in data science in general.

To complement our quantitative evaluation it would naturally be of interest to evaluate the performance of treemapping algorithms in various usage scenarios through user studies. The two metrics we use for visual quality and stability are both perceptually salient according to studies performed in previous work. However, a study that evaluates the combination of and the trade-offs between visual quality and stability could deliver important insights as to where on the Pareto-front an optimal treemapping algorithm should lie.

Finally, our evaluation currently does not measure the run-time and correspondingly the scalability of the algorithms used in our experiments. Our implementations are not (equally) optimized and hence a fair comparison is currently impossible. Scalability is clearly an important factor in online usage scenarios, and we hope to be able to complement our current set of implementations with optimized versions in the near future.

\mypar{Acknowledgments}
The Netherlands Organisation for Scientific Research (NWO) is supporting M. Sondag and B. Speckmann under project no. 639.023.208, and K. Verbeek under project no. 639.021.541. This study was also financed in part by CAPES (Finance Code 001) and CNPq (Process 308851/2015-3).

\bibliographystyle{eg-alpha-doi}
\bibliography{treemaps.bib}

\newcommand{\etalchar}[1]{$^{#1}$}
\begin{thebibliography}{\uppercase{VvWvdL06}}

\bibitem[BDL05]{balzer05b}
\textsc{Balzer M., Deussen O., Lewerentz C.}:
\newblock {V}oronoi treemaps for the visualization of software metrics.
\newblock In \emph{Proc. ACM Symposium on Software Visualization} (2005),
  pp.~165--172.

\bibitem[BHvW00]{sqr}
\textsc{Bruls M., Huizing K., van Wijk J.~J.}:
\newblock Squarified treemaps.
\newblock In \emph{Proc. Data Visualization} (2000), pp.~33--42.

\bibitem[Bie87]{biederman87}
\textsc{Biederman I.}:
\newblock Recognition-by-components: A theory of human image understanding.
\newblock \emph{Psychological Review 94}, 2 (1987), 115--147.

\bibitem[Bis06]{bishop06}
\textsc{Bishop C.}:
\newblock \emph{Pattern Recognition and Machine Learning}.
\newblock Springer, 2006.

\bibitem[BO73]{boorman73}
\textsc{Boorman S.~A., Oliviera D.~C.}:
\newblock Metrics on spaces of finite trees.
\newblock In \emph{Journal of Mathematical Psychology} (1973), vol.~10,
  pp.~26--59.

\bibitem[BSW02]{bederson02}
\textsc{Bederson B.~B., Shneiderman B., Wattenberg M.}:
\newblock Ordered and quantum treemaps: Making effective use of {2D} space to
  display hierarchies.
\newblock \emph{ACM Transactions on Graphics 21}, 4 (2002), 833--854.

\bibitem[CDY17]{Chen2017}
\textsc{Chen Y., Du X., Yuan X.}:
\newblock Ordered small multiple treemaps for visualizing time-varying
  hierarchical pesticide residue data.
\newblock \emph{Visual Computer 33}, 6 (2017), 1073--1084.

\bibitem[CSP{\etalchar{*}}06]{TimeTree}
\textsc{Card S., Suh B., Pendleton B.~A., Heer B., Bodnar J.~W.}:
\newblock Time tree: Exploring time changing hierarchies.
\newblock In \emph{Proc. Symposium on Visual Analytics Science and Technology}
  (2006), pp.~3--10.

\bibitem[dBSvdW14]{deberg14}
\textsc{de~Berg M., Speckmann B., van~der Weele V.}:
\newblock Treemaps with bounded aspect ratio.
\newblock \emph{Computational Geometry 47}, 6 (2014), 683 -- 693.

\bibitem[EMK{\etalchar{*}}19]{espadoto19}
\textsc{Espadoto M., Martins R., Kerren A., Hirata N., Telea A.}:
\newblock Towards a quantitative survey of dimension reduction techniques.
\newblock \emph{IEEE Transactions on Visualization and Computer Graphics}
  (2019).

\bibitem[EMSV12]{eppstein2009area}
\textsc{Eppstein D., Mumford E., Speckmann B., Verbeek K.}:
\newblock Area-universal and constrained rectangular layouts.
\newblock \emph{SIAM Journal on Computing 41}, 3 (2012), 537--564.

\bibitem[Eng05]{engdahl}
\textsc{Engdahl B.}:
\newblock Ordered and unordered treemap algorithms and their applications on
  handheld devices.
\newblock \emph{Master's degree project, Department of Numerical Analysis and
  Computer Science, Stockholm Royal Institute of Technology, SE-100 44} (2005).

\bibitem[GGPPS13]{StemView}
\textsc{Guerra-G{\'o}mez J., Pack M., Plaisant C., Shneiderman B.}:
\newblock Visualizing change over time using dynamic hierarchies:
  {T}ree{V}ersity2 and the {S}tem{V}iew.
\newblock \emph{IEEE Transactions on Visualization and Computer Graphics 19},
  12 (2013), 2566--2575.

\bibitem[Hah15]{hahn2015comparing}
\textsc{Hahn S.}:
\newblock Comparing the layout stability of treemap algorithms.
\newblock \emph{Proc. HPI research school on service-oriented systems
  engineering 95} (2015), 71--79.

\bibitem[HBD17]{Hahn2017}
\textsc{Hahn S., Bethge J., D{\"o}llner J.}:
\newblock Relative direction change -- a topology-based metric for layout
  stability in treemaps.
\newblock In \emph{Proc. International Conference on Information Visualization
  Theory and Applications} (2017), pp.~88--95.

\bibitem[HK16]{harper2016movielens}
\textsc{Harper F.~M., Konstan J.~A.}:
\newblock The movielens datasets: History and context.
\newblock \emph{ACM Transactions on Interactive Intelligent Systems 5}, 4
  (2016), 19.

\bibitem[HTMD14]{hahn10}
\textsc{Hahn S., Tr{\"u}mper J., Moritz D., D{\"o}llner J.}:
\newblock Visualization of varying hierarchies by stable layout of voronoi
  treemaps.
\newblock In \emph{Proc. International Conference on Information Visualization
  Theory and Applications} (2014), pp.~50--58.

\bibitem[KHA10]{Kong2010}
\textsc{Kong N., Heer J., Agrawala M.}:
\newblock Perceptual guidelines for creating rectangular treemaps.
\newblock \emph{IEEE Transactions on Visualization and Computer Graphics 16}, 6
  (2010), 990--998.

\bibitem[KW19]{kopp2019temporal}
\textsc{K{\"o}pp W., Weinkauf T.}:
\newblock Temporal treemaps: Static visualization of evolving trees.
\newblock \emph{IEEE Transactions on Visualization and Computer Graphics 25}, 1
  (2019), 534--543.

\bibitem[KY15]{kuhner14}
\textsc{Kuhner M.~K., Yamato J.}:
\newblock Practical performance of tree comparison metrics.
\newblock \emph{Systematic Biology 64}, 2 (2015), 205--214.

\bibitem[LFH{\etalchar{*}}17]{lu2017golden}
\textsc{Lu L., Fan S., Huang M., Huang W., Yang R.}:
\newblock Golden rectangle treemap.
\newblock \emph{Journal of Physics: Conference Series 787}, 1 (2017).

\bibitem[LWM{\etalchar{*}}17]{lukasczyk2017nested}
\textsc{Lukasczyk J., Weber G., Maciejewski R., Garth C., Leitte H.}:
\newblock Nested tracking graphs.
\newblock \emph{Computer Graphics Forum 36}, 3 (2017), 12--22.

\bibitem[{Mee}16]{urlmeertens}
\textsc{{Meertens Instituut, KNAW}}:
\newblock Nederlandse voornamenbank.
\newblock \url{https://www.meertens.knaw.nl/nvb}, accessed 30-05-2016.

\bibitem[NA07]{nagamochi2007approximation}
\textsc{Nagamochi H., Abe Y.}:
\newblock An approximation algorithm for dissecting a rectangle into rectangles
  with specified areas.
\newblock \emph{Discrete Applied Mathematics 155}, 4 (2007), 523--537.

\bibitem[S{\etalchar{*}}94]{shi1994good}
\textsc{Shi J., et~al.}:
\newblock Good features to track.
\newblock In \emph{Proc. IEEE Conference on Computer Vision and Pattern
  Recognition} (1994), pp.~593--600.

\bibitem[Shn92]{shneiderman92}
\textsc{Shneiderman B.}:
\newblock Tree visualization with tree-maps: a {2D} space-filling approach.
\newblock \emph{ACM Transactions on Graphics 11}, 1 (1992), 92--99.

\bibitem[SMBWL14]{smith2014towards}
\textsc{Smith-Miles K., Baatar D., Wreford B., Lewis R.}:
\newblock Towards objective measures of algorithm performance across instance
  space.
\newblock \emph{Computers \& Operations Research 45} (2014), 12--24.

\bibitem[SSV18]{sondag17}
\textsc{Sondag M., Speckmann B., Verbeek K.}:
\newblock Stable treemaps via local moves.
\newblock \emph{IEEE Transactions on Visualization and Computer Graphics 24}, 1
  (2018), 729--738.

\bibitem[SW01]{ordered}
\textsc{Shneiderman B., Wattenberg M.}:
\newblock Ordered treemap layouts.
\newblock In \emph{Proc. IEEE Symposium on Information Visualization} (2001),
  pp.~73--78.

\bibitem[SWD18]{Scheibel2018}
\textsc{Scheibel W., Weyand C., D{\"o}llner J.}:
\newblock {EvoCells} -- a treemap layout algorithm for evolving tree data.
\newblock In \emph{Proc. International Conference on Information Visualization
  Theory and Applications} (2018), pp.~273--280.

\bibitem[Sze10]{szeliski10}
\textsc{Szeliski R.}:
\newblock \emph{Computer Vision: Algorithms and Applications}.
\newblock Springer, 2010.

\bibitem[TC13]{hilbert_moore}
\textsc{Tak S., Cockburn A.}:
\newblock Enhanced spatial stability with {H}ilbert and {M}oore treemaps.
\newblock \emph{IEEE Transactions on Visualization and Computer Graphics 19}, 1
  (2013), 141--148.

\bibitem[Tel06]{tablelens}
\textsc{Telea A.}:
\newblock Combining extended table lens and treemap techniques for visualizing
  tabular data.
\newblock In \emph{Proc. VGTC Conference on Visualization} (2006),
  pp.~120--127.

\bibitem[TM07]{tuytelaars07}
\textsc{Tuytelaars T., Mikolajczyk K.}:
\newblock Local invariant feature detectors: A survey.
\newblock \emph{Foundations and Trends in Computer Graphics and Vision 3}, 3
  (2007), 177--280.

\bibitem[TS07]{spiral}
\textsc{Tu Y., Shen H.}:
\newblock Visualizing changes of hierarchical data using treemaps.
\newblock \emph{IEEE Transactions on Visualization and Computer Graphics 13}, 6
  (2007), 1286--1293.

\bibitem[URLa]{URLscitools}
{Scitools}.
\newblock \url{https://scitools.com}.

\bibitem[URLb]{URLTreemaps}
{Treemap resources}.
\newblock
  \url{https://eduardovernier.github.io/dynamic-treemap-resources-eurovis}.

\bibitem[URL17]{URLComtrade}
{UN Comtrade Database}.
\newblock \url{https://comtrade.un.org}, accessed 15-02-2017.

\bibitem[URL18a]{URLatp}
{ATP Tennis Rankings}.
\newblock \url{https://github.com/JeffSackmann/tennis_atp}, accessed
  03-07-2018.

\bibitem[URL18b]{URLearth}
{USGS Earthquakes}.
\newblock \url{https://earthquake.usgs.gov/earthquakes/browse/stats.php},
  accessed 03-07-2018.

\bibitem[URL18c]{URLWorldbank}
{Worldbank indicators}.
\newblock \url{https://data.worldbank.org/indicator/}, accessed 04-07-2018.

\bibitem[URL18d]{URLmdb}
{The Movie Database}.
\newblock \url{www.themoviedb.org}, accessed 10-02-2018.

\bibitem[URL18e]{URLGit}
{Github}.
\newblock \url{https://github.com}, accessed 16-07-2018.

\bibitem[VCT18a]{vernier18software}
\textsc{Vernier E., Comba J., Telea A.}:
\newblock Quantitative comparison of dynamic treemaps for software evolution
  visualization.
\newblock In \emph{IEEE Conference on Software Visualization} (2018),
  pp.~96--106.

\bibitem[VCT18b]{vernier18git}
\textsc{Vernier E., Comba J., Telea A.}:
\newblock A stable greedy insertion treemap algorithm for software evolution
  visualization.
\newblock In \emph{IEEE Conference on Graphics, Patterns and Images} (2018),
  pp.~158--165.

\bibitem[vHH17]{hees17}
\textsc{van Hees R., Hage J.}:
\newblock Stable and predictable {V}oronoi treemaps for software quality
  monitoring.
\newblock \emph{Information and Software Technology 87} (2017), 242 -- 258.

\bibitem[VvWvdL06]{vliegen}
\textsc{Vliegen R., van Wijk J.~J., van~der Linden E.~J.}:
\newblock Visualizing business data with generalized treemaps.
\newblock \emph{IEEE Transactions on Visualization and Computer Graphics 12}, 5
  (2006), 789--796.

\bibitem[Wat05]{jigsaw}
\textsc{Wattenberg M.}:
\newblock A note on space-filling visualizations and space-filling curves.
\newblock In \emph{Proc. IEEE Symposium on Information Visualization} (2005),
  pp.~181--186.

\bibitem[ZCYT17]{Zhou2017}
\textsc{Zhou M., Cheng Y., Ye N., Tian J.}:
\newblock Effectiveness and efficiency of using different types of rectangular
  treemap as diagrams in cartography.
\newblock In \emph{International Cartographic Conference} (2017), pp.~187--206.

\end{thebibliography}

\end{document}